\documentclass[showpacs,aps,graphicx,preprint]{revtex4}%
\usepackage{graphicx}
\begin{document}

\title{Entanglement  concentration for concatenated
Greenberger-Horne-Zeiglinger state with feasible linear optics}

\author{Yu-Bo Sheng,$^{1}$\footnote{Corresponding author:
Email address: shengyb@njupt.edu.cn} Chang-Cheng Qu,$^1$ Lan Zhou$^{1,2}$ }
\address{$^1$Key Lab of Broadband Wireless Communication and Sensor Network
 Technology, Nanjing University of Posts and Telecommunications, Ministry of
 Education, Nanjing, 210003, China\\
 $^2$College of Mathematics \& Physics, Nanjing University of Posts and Telecommunications, Nanjing,
210003, China\\}
\date{\today }
\begin{abstract}
The concatenated Greenberger-Horne-Zeiglinger (C-GHZ) state which is a new type of logic-qubit entanglement  has attracted a lot of attentions recently.  We present a feasible  entanglement concentration protocol (ECP) for logic-qubit
entanglement. This ECP is based on the linear optics, and it does not know the initial coefficients of the less-entangled C-GHZ state. This protocol can be extended to arbitrary C-GHZ state. This protocol may be useful in future quantum information processing tasks. 
\end{abstract}
 \maketitle
\section{Introduction}
Quantum entanglement plays a very important role in the field of current quantum information processing\cite{Entanglement1,Entanglement2}, such as quantum key distribution\cite{Distribution1,Distribution2}, quantum teleportation\cite{Teleportation}, quantum secure direct communication\cite{Direct1,Direct2} and some other important quantum communication protocols\cite{Proc4,Proc5,Proc6}, which are all require the maximal entanglement to set up a secure quantum channel. However, entanglement is generally fragile and the  noise in environment always exists. Owing to the interaction with the environment, channel noise cannot be avoided during transmission and storage, which will make the maximally entangled states degrade to mixed states or pure less-entangled states. In practical application, such mixed states or less-entangled states may decrease and one cannot ultimately set up the high quality quantum entanglement channel successfully \cite{Su}. Therefore, we need to recover the mixed state or the pure less-entangled state into the maximally entangled state before starting quantum communication.

The method of converting the mixed entangled state into the maximally entangled states is called entanglement purification protocol (EPP), which has been widely discussed in recent years\cite{Ep0,Ep1,Ep2,Ep3,Ep4,Ep5,Ep6,Ep7,Ep8}. Entanglement concentration protocol (ECP), which will be discussed in detail here, is a powerful approach to extract the maximally entangled state from the pure partially entangled state with local operation and classical communication (LOCC)\cite{Locc1,Locc2,Locc3,Locc4,Locc5,Locc6,Locc7,Locc8,Locc9,Locc10,Locc11,Locc12,Locc13,Locc14,Locc15,Locc16,Locc17,Locc18,Locc19,Locc20,Locc21,Locc22,Locc23,Locc24,
Locc25,Locc26,Locc27,Locc28,Locc29,Locc30,Locc31,Locc32,Locc33,Locc34,Locc35,Locc36,Locc37,Locc38,Locc39,Locc40,Locc41,Locc42,Locc43,Qu,panjun}. For example,   Bennett \emph{et al.} put forward the first ECP which is called the Schmidt projection protocol in 1996 \cite{Locc1}. Bose \emph{et al.} proposed the ECP based on entanglement swapping\cite{Locc2}. In 2001, Zhao \emph{et al.} and Yamamoto \emph{et al.} independently put forward two similar ECPs with linear optics \cite{Locc4,Locc5}. In 2008, Sheng \emph{et al.} proposed ECP with higher success efficiency which is based on cross-Kerr nonlinearity\cite{Locc6}. In 2012, by making use of local operation and classical communications, Sheng \emph{et al.} and Deng  independently proposed two  ECPs assisted with single photons which made concentration process more simple and useful in practical applications \cite{Locc7,Locc8}. Subsequently, their protocols were widely used to deal with the different kinds of entanglement concentration, such as the multi-photon W state entanglement concentration \cite{Locc10,Locc11,Locc37}, cluster state entanglement concentration \cite{Locc12}, $\chi$-type entanglement concentration \cite{Locc25}, hyperentanglement concentration\cite{Locc15,Locc16,Locc32,Locc38}, and so on.

In the practical quantum information processing, the logic qubit, which encodes many physical qubits in a logic qubit\cite{Gh1,Gh2,Gh3,Gh4}, has been widely discussed. In 2011, Fr\"{o}wis et al. firstly put forward the concatenated Greenberger-Horne-Zeilinger (C-GHZ) state, which is a promising logic-qubit entanglement for quantum information processing \cite{Gh5}. The form of the C-GHZ state can be described as $|\Psi\rangle_{N,m}=\frac{1}{\sqrt{2}}(|GHZ^{+}_{m}\rangle^{\otimes N}+|GHZ^{-}_{m}\rangle^{\otimes N})$, where $|GHZ^{\pm}_{m}\rangle=\frac{1}{\sqrt{2}}(|0\rangle^{\otimes m}\pm|1\rangle^{\otimes m})$. The $N$ is the number of logical qubits, and the $m$ is the number of physical qubit encoded in each logical qubit. In 2014, Lu \emph{et al.} described the first experiment for realizing such C-GHZ state with linear optics \cite{Gh6}. Both the theory and experiment showed that the C-GHZ is useful in future quantum communication. Recently, the approaches of discrimination of the C-GHZ state were proposed\cite{G1,G2,G3}.

As the C-GHZ state has its potential application in quantum information processing,  we will describe an efficient ECP for the less-entangled
C-GHZ state with the help of linear optics.  After performing this ECP, we can ultimately obtain the maximally C-GHZ state with some success probability.
Moreover, this protocol exploits the polarization beam splitters (PBSs), and have-wave plates (HWPs) to complete the task. It is feasible in current technology. This paper is organized as follows. In Sec. II, we first describe this protocol with a simple case with $m=N=2$. In Sec. III, we explain the ECP for a more complicated case that $m=3$ and $N=2$. We show that if each logic qubit is the multiphoton GHZ state, this protocol can also work. In Sec. IV, we extend this ECP for arbitrary C-GHZ. Finally, In Sec. V, we will present a discussion and conclusion.

\section{ECP for C-GHZ state with m=N=2}
  The less-entangled C-GHZ state can be written as
\begin{eqnarray}
|\Psi\rangle_{N,m}=\alpha|GHZ^{+}_{m}\rangle^{\otimes N}+\beta|GHZ^{-}_{m}\rangle^{\otimes N},
\end{eqnarray}
here $|GHZ^{\pm}_{m}\rangle^{\otimes N}=\frac{1}{\sqrt{2}}(|H\rangle^{\otimes m}\pm|V\rangle^{\otimes m})$,
and $|\alpha|^{2}+|\beta|^{2}=1.$ $|H\rangle$ is the horizontal polarization  and $|V\rangle$ is the vertical polarization of the photon, respectively. Our aim is to concentrate the less-entangled C-GHZ state to the maximally C-GHZ state as
\begin{eqnarray}
|\Psi\rangle'_{N,m}=\frac{1}{\sqrt{2}}(|GHZ^{+}_{m}\rangle^{\otimes N}+|GHZ^{-}_{m}\rangle^{\otimes N}).
\end{eqnarray}

In this section, we will discuss our ECP in the first group with a simple case that concentrate the  less-entangled C-GHZ state with $m=N=2$.
If $m=2$, it shows that each logic qubit is a Bell state.
Such C-GHZ state can be written as
\begin{eqnarray}
|\Psi\rangle_{s}&=&\alpha|GHZ^{+}_{2}\rangle^{\otimes 2}+\beta|GHZ^{-}_{2}\rangle^{\otimes 2}\nonumber\\
&=&\alpha[\frac{1}{\sqrt{2}}(|H\rangle|H\rangle+|V\rangle|V\rangle)\frac{1}{\sqrt{2}}(|H\rangle|H\rangle+|V\rangle|V\rangle)]\nonumber\\
&&+\beta[\frac{1}{\sqrt{2}}(|H\rangle|H\rangle-|V\rangle|V\rangle)\frac{1}{\sqrt{2}}(|H\rangle|H\rangle-|V\rangle|V\rangle)]\nonumber\\ &=&\frac{1}{2}\alpha(|H\rangle|H\rangle+|V\rangle|V\rangle)(|H\rangle|H\rangle+|V\rangle|V\rangle)\nonumber\\
&&+\frac{1}{2}\beta(|H\rangle|H\rangle-|V\rangle|V\rangle)(|H\rangle|H\rangle-|V\rangle|V\rangle).\label{cghz1}
\end{eqnarray}
\begin{figure}[!h]
\begin{center}
\includegraphics[width=10cm,angle=0]{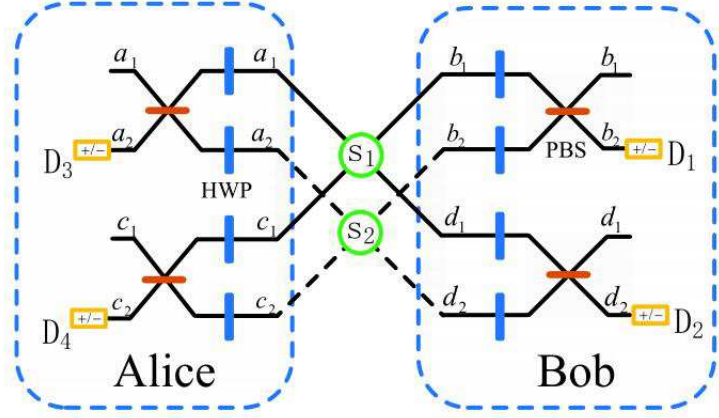}
\caption{Schematic diagram of the entanglement concentration. $|\Psi\rangle_{s1}$ and  $|\Psi\rangle_{s2}$ are the two copies of less-entangled C-GHZ state. $S_{1}$ and $S_{2}$ are the entanglement sources. The HWP represents the Hadamard operation which makes the $|H\rangle\rightarrow\frac{1}{\sqrt{2}}(|H\rangle+|V\rangle)$ and $|V\rangle\rightarrow\frac{1}{\sqrt{2}}(|H\rangle-|V\rangle)$. The PBSs are polarized beam splitters. $D_{1}$ and $D_{2}$ represent the photon number detectors. $|\pm\rangle$ denote the measurement of output photos in the basis $|\pm\rangle=\frac{1}{\sqrt{2}}(|H\rangle+|V\rangle)$.}\label{f1}
\end{center}
\end{figure}

Fig. \ref{f1} is the schematic drawing of our entanglement concentration. We first choose two copies of less-entangled C-GHZ states  $|\Psi\rangle_{s1}$ and $|\Psi\rangle_{s2}$ which can  be  written as
\begin{eqnarray}
|\Psi\rangle_{s_{1}}&=&\frac{1}{2}\alpha(|H\rangle_{a_{1}}|H\rangle_{c_{1}}+|V\rangle_{a_{1}}|V\rangle_{c_{1}})(|H\rangle_{b_{1}}|H\rangle_{d_{1}}+|V\rangle_{b_{1}}|V\rangle_{d_{1}})\nonumber\\
&+&\frac{1}{2}\beta(|H\rangle_{a_{1}}|H\rangle_{c_{1}}-|V\rangle_{a_{1}}|V\rangle_{c_{1}})(|H\rangle_{b_{1}}|H\rangle_{d_{1}}-|V\rangle_{b_{1}}|V\rangle_{d_{1}}),\label{4}\\
|\Psi\rangle_{s_{2}}&=&\frac{1}{2}\beta(|H\rangle_{a_{2}}|H\rangle_{c_{2}}+|V\rangle_{a_{2}}|V\rangle_{c_{2}})(|H\rangle_{b_{2}}|H\rangle_{d_{2}}+|V\rangle_{b_{2}}|V\rangle_{d_{2}})\nonumber\\
&+&\frac{1}{2}\alpha(|H\rangle_{a_{2}}|H\rangle_{c_{2}}-|V\rangle_{a_{2}}|V\rangle_{c_{2}})(|H\rangle_{b_{2}}|H\rangle_{d_{2}}-|V\rangle_{b_{2}}|V\rangle_{d_{2}}).
\end{eqnarray}
 It is shown that the photons $a_{1}$, $c_{1}$, $a_{2}$ and $c_{2}$ are sent to Alice and the photons $b_{1}$, $d_{1}$, $b_{2}$ and $d_{2}$ are sent to Bob, respectively.  Before  concentration, we should convert the state $|\Psi\rangle_{s2}$ into $|\Psi\rangle'_{s2}$ with the help of Hadamard operation and single qubit rotation. The former of $|\Psi\rangle'_{s2}$ can be written as
\begin{eqnarray}
|\Psi\rangle'_{s_{2}}&=&\frac{1}{2}\beta(|H\rangle_{a_{2}}|H\rangle_{c_{2}}+|V\rangle_{a_{2}}|V\rangle_{c_{2}})(|H\rangle_{b_{2}}|H\rangle_{d_{2}}+|V\rangle_{b_{2}}|V\rangle_{d_{2}})\nonumber\\
&+&\frac{1}{2}\alpha(|H\rangle_{a_{2}}|H\rangle_{c_{2}}-|V\rangle_{a_{2}}|V\rangle_{c_{2}})(|H\rangle_{b_{2}}|H\rangle_{d_{2}}-|V\rangle_{b_{2}}|V\rangle_{d_{2}}).
\end{eqnarray}

Firstly, the parties Alice and Bob let the photons $|\Psi\rangle_{s1}$ and $|\Psi\rangle'_{s2}$ pass through the half-wave plates (HWP), which will make $|H\rangle\rightarrow\frac{1}{\sqrt{2}}(|H\rangle+|V\rangle)$, and $|V\rangle\rightarrow\frac{1}{\sqrt{2}}(|H\rangle-|V\rangle)$.
After the HWP operations, the state $|\Psi\rangle_{s1}\otimes|\Psi\rangle'_{s2}$ evolves to
\begin{eqnarray}
|\Psi\rangle_{s_{3}}&=&\frac{1}{4}\alpha^{2}(|H\rangle_{a_{1}}|H\rangle_{c_{1}}+|V\rangle_{a_{1}}|V\rangle_{c_{1}})(|H\rangle_{b_{1}}|H\rangle_{d_{1}}+|V\rangle_{b_{1}}|V\rangle_{d_{1}})\nonumber\\
&&(|H\rangle_{a_{2}}|V\rangle_{c_{2}}+|V\rangle_{a_{2}}|H\rangle_{c_{2}})(|H\rangle_{b_{2}}|V\rangle_{d_{2}}+|V\rangle_{b_{2}}|H\rangle_{d_{2}})\nonumber\\
&+&\frac{1}{4}\beta^{2}(|H\rangle_{a_{1}}|V\rangle_{c_{1}}+|V\rangle_{a_{1}}|H\rangle_{c_{1}})(|H\rangle_{b_{1}}|V\rangle_{d_{1}}+|V\rangle_{b_{1}}|H\rangle_{d_{1}})\nonumber\\
&&(|H\rangle_{a_{2}}|H\rangle_{c_{2}}+|V\rangle_{a_{2}}|V\rangle_{c_{2}})(|H\rangle_{b_{2}}|H\rangle_{d_{2}}+|V\rangle_{b_{2}}|V\rangle_{d_{2}})\nonumber\\
&+&\frac{1}{4}\alpha\beta[(|H\rangle_{a_{1}}|H\rangle_{c_{1}}+|V\rangle_{a_{1}}|V\rangle_{c_{1}})(|H\rangle_{b_{1}}|H\rangle_{d_{1}}+|V\rangle_{b_{1}}|V\rangle_{d_{1}})\nonumber\\
&&(|H\rangle_{a_{2}}|H\rangle_{c_{2}}+|V\rangle_{a_{2}}|V\rangle_{c_{2}})(|H\rangle_{b_{2}}|H\rangle_{d_{2}}+|V\rangle_{b_{2}}|V\rangle_{d_{2}})\nonumber\\
&+&(|H\rangle_{a_{1}}|V\rangle_{c_{1}}+|V\rangle_{a_{1}}|H\rangle_{c_{1}})(|H\rangle_{b_{1}}|V\rangle_{d_{1}}+|V\rangle_{b_{1}}|H\rangle_{d_{1}})\nonumber\\
&&(|H\rangle_{a_{2}}|V\rangle_{c_{2}}+|V\rangle_{a_{2}}|H\rangle_{c_{2}})(|H\rangle_{b_{2}}|V\rangle_{d_{2}}+|V\rangle_{b_{2}}|H\rangle_{d_{2}})].\label{evolve1}
\end{eqnarray}
Subsequently, as shown in Fig. \ref{f1}, the parties Alice and Bob let the photons  $a_{1}$ and $a_{2}$, $c_{1}$ and $c_{2}$, $b_{1}$ and $b_{2}$ and $d_{1}$ and $d_{2}$ pass through the four PBSs which are optical devices to transmit horizontally polarized light and reflecting light of vertical polarization.
If all the output modes of the PBSs only contain one photon, the state in Eq. (\ref{evolve1}) can be written as
\begin{eqnarray}
&\rightarrow&\frac{1}{2\sqrt{2}}(|H\rangle_{a_{1}}|H\rangle_{c_{1}}|H\rangle_{a_{2}}|H\rangle_{c_{2}}+|V\rangle_{a_{1}}|V\rangle_{c_{1}}|V\rangle_{a_{2}}|V\rangle_{c_{2}})\nonumber\\
&&(|H\rangle_{b_{1}}|H\rangle_{d_{1}}|H\rangle_{b_{2}}|H\rangle_{d_{2}}+|V\rangle_{b_{1}}|V\rangle_{d_{1}}|V\rangle_{b_{2}}|V\rangle_{d_{2}})\nonumber\\
&+&(|H\rangle_{a_{1}}|V\rangle_{c_{1}}|H\rangle_{a_{2}}|V\rangle_{c_{2}}+|V\rangle_{a_{1}}|H\rangle_{c_{1}}|V\rangle_{a_{2}}|H\rangle_{c_{2}})\nonumber\\
&&(|H\rangle_{b_{1}}|V\rangle_{d_{1}}|H\rangle_{b_{2}}|V\rangle_{d_{2}}+|V\rangle_{b_{1}}|H\rangle_{d_{1}}|V\rangle_{b_{2}}|H\rangle_{d_{2}}).\nonumber\\
\end{eqnarray}

And then, the photons detected by the four photon detectors $D_{1}D_{2}D_{3}D_{4}$ in the base $|\pm\rangle$. Only if each photon detector detects a single photon, it means that the ECP is successful. If the polarization measurement is $|++\rangle_{a_{2}c_{2}}$ or $|--\rangle_{a_{2}c_{2}}$, and $|++\rangle_{b_{2}d_{2}}$ or $|--\rangle_{b_{2}d_{2}}$, they will obtain the state
\begin{eqnarray}
|\Psi\rangle_{s_{4}}&=&\frac{1}{\sqrt{2}}[\frac{1}{\sqrt{2}}(|H\rangle_{a_{1}}|H\rangle_{c_{1}}+|V\rangle_{a_{1}}|V\rangle_{c_{1}})\frac{1}{\sqrt{2}}(|H\rangle_{b_{1}}|H\rangle_{d_{1}}+|V\rangle_{b_{1}}|V\rangle_{d_{1}})\nonumber\\
&+&\frac{1}{\sqrt{2}}(|H\rangle_{a_{1}}|V\rangle_{c_{1}}+|V\rangle_{a_{1}}|H\rangle_{c_{1}})\frac{1}{\sqrt{2}}(|H\rangle_{b_{1}}|V\rangle_{d_{1}}+|V\rangle_{b_{1}}|H\rangle_{d_{1}})].\label{rt}
\end{eqnarray}
If the polarization measurement results are  $|+-\rangle_{a_{2}c_{2}}$ or $|+-\rangle_{b_{2}d_{2}}$, and $|-+\rangle_{a_{2}c_{2}}$ or $|-+\rangle_{b_{2}d_{2}}$,
they will obtain the state
\begin{eqnarray}
|\Psi\rangle'_{s_{4}}&=&\frac{1}{\sqrt{2}}[\frac{1}{\sqrt{2}}(|H\rangle_{a_{1}}|H\rangle_{c_{1}}-|V\rangle_{a_{1}}|V\rangle_{c_{1}})\frac{1}{\sqrt{2}}(|H\rangle_{b_{1}}|H\rangle_{d_{1}}-|V\rangle_{b_{1}}|V\rangle_{d_{1}})\nonumber\\
&+&\frac{1}{\sqrt{2}}(|H\rangle_{a_{1}}|V\rangle_{c_{1}}-|V\rangle_{a_{1}}|H\rangle_{c_{1}})\frac{1}{\sqrt{2}}(|H\rangle_{b_{1}}|V\rangle_{d_{1}}-|V\rangle_{b_{1}}|H\rangle_{d_{1}})].\label{rt1}
\end{eqnarray}
By performing the phase-file operation  they can also finally obtain the state in Eq. (\ref{rt}). In order to get the maximally entangled C-GHZ state, they have to perform Hadamard operations on each photons to convert Eq. (\ref{rt}) to the maximally C-GHZ state as
\begin{eqnarray}
|\Psi\rangle_{s_{5}}&=&\frac{1}{\sqrt{2}}[\frac{1}{\sqrt{2}}(|H\rangle_{a_{1}}|H\rangle_{c_{1}}+|V\rangle_{a_{1}}|V\rangle_{c_{1}})\frac{1}{\sqrt{2}}(|H\rangle_{b_{1}}|H\rangle_{d_{1}}+|V\rangle_{b_{2}}|V\rangle_{d_{1}})\nonumber\\
&+&\frac{1}{\sqrt{2}}(|H\rangle_{a_{1}}|H\rangle_{c_{1}}-|V\rangle_{a_{1}}|V\rangle_{c_{1}})\frac{1}{\sqrt{2}}(|H\rangle_{b_{1}}|H\rangle_{d_{1}}-|V\rangle_{b_{1}}|V\rangle_{d_{1}})].\label{ss}
\end{eqnarray}
 We get the maximally C-GHZ state with the total success probability of $\frac{1}{2}|\alpha\beta|^{2}$.

\section{The ECP for C-GHZ state with $m=3$, $N=2$}

In this section, we discuss the ECP for the C-GHZ state with $m=3$ and $N=2$. The C-GHZ state with $m=3$ and $N=2$ can be written as
\begin{eqnarray}
|\Gamma\rangle_{s_{1}}&=&\alpha[\frac{1}{\sqrt{2}}(|H\rangle_{a_{1}}|H\rangle_{c_{1}}|H\rangle_{t_{1}}+|V\rangle_{a_{1}}|V\rangle_{c_{1}}|V\rangle_{t_{1}})\frac{1}{\sqrt{2}}(|H\rangle_{b_{1}}|H\rangle_{d_{1}}|H\rangle_{h_{1}}+|V\rangle_{b_{1}}|V\rangle_{d_{1}}|V\rangle_{h_{1}})]\nonumber\\
&+&\beta[\frac{1}{\sqrt{2}}(|H\rangle_{a_{1}}|H\rangle_{c_{1}}|H\rangle_{t_{1}}-|V\rangle_{a_{1}}|V\rangle_{c_{1}}|V\rangle_{t_{1}})\frac{1}{\sqrt{2}}(|H\rangle_{b_{1}}|H\rangle_{d_{1}}|H\rangle_{h_{1}}-|V\rangle_{b_{1}}|V\rangle_{d_{1}}|V\rangle_{h_{1}})]\nonumber\\ &=&\frac{1}{2}\alpha(|H\rangle_{a_{1}}|H\rangle_{c_{1}}|H\rangle_{t_{1}}+|V\rangle_{a_{1}}|V\rangle_{c_{1}}|V\rangle_{t_{1}})(|H\rangle_{b_{1}}|H\rangle_{d_{1}}|H\rangle_{h_{1}}+|V\rangle_{b_{1}}|V\rangle_{d_{1}}|V\rangle_{h_{1}})\nonumber\\
&+&\frac{1}{2}\beta(|H\rangle_{a_{1}}|H\rangle_{c_{1}}|H\rangle_{t_{1}}-|V\rangle_{a_{1}}|V\rangle_{c_{1}}|V\rangle_{t_{1}})(|H\rangle_{b_{1}}|H\rangle_{d_{1}}|H\rangle_{h_{1}}-|V\rangle_{b_{1}}|V\rangle_{d_{1}}|V\rangle_{h_{1}}).\label{11}
\end{eqnarray}
The copy of $|\Gamma\rangle_{s_{1}}$ is
In order to perform the concentration, they also select two copies of the less-entangled state $|\Gamma\rangle_{s_{1}}$ and $|T\rangle'_{s_{1}}$ as shown in Eq. (\ref{11}).
Similar to the ECP we discussed in the previous section, we firstly convert the initial state $|T\rangle'_{s_{1}}$ into
\begin{eqnarray}
|T\rangle_{s_{2}}&=&\frac{1}{2}\beta(|H\rangle_{a_{2}}|H\rangle_{c_{2}}|H\rangle_{t_{2}}+|V\rangle_{a_{2}}|V\rangle_{c_{2}}|V\rangle_{t_{2}})(|H\rangle_{b_{2}}|H\rangle_{d_{2}}|H\rangle_{h_{2}}+|V\rangle_{b_{1}}|V\rangle_{d_{2}}|V\rangle_{h_{2}})\nonumber\\
&+&\frac{1}{2}\alpha(|H\rangle_{a_{2}}|H\rangle_{c_{2}}|H\rangle_{t_{2}}-|V\rangle_{a_{2}}|V\rangle_{c_{2}}|V\rangle_{t_{2}})(|H\rangle_{b_{2}}|H\rangle_{d_{2}}|H\rangle_{h_{2}}-|V\rangle_{b_{2}}|V\rangle_{d_{1}}|V\rangle_{h_{2}}).
\end{eqnarray}
\begin{figure}[!h]
\begin{center}
\includegraphics[width=8cm,angle=0]{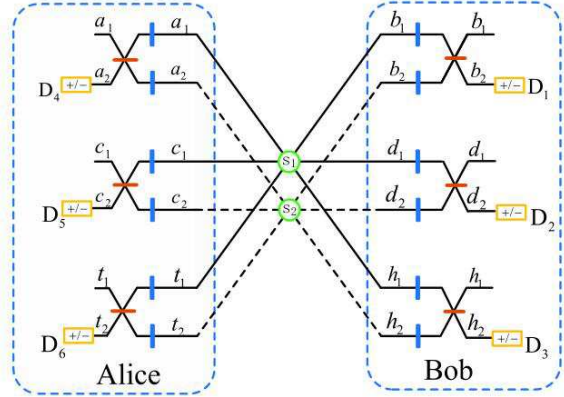}
\caption{Schematic diagram of the entanglement concentration about the C-GHZ state with $m=3$, $N=2$.}\label{f2}
\end{center}
\end{figure}

As shown ing Fig. \ref{f2}, the photons in spatial modes $a_{1}c_{1}t_{1}$ and $a_{2}c_{2}t_{2}$ are distributed to Alice and the photons in spatial modes $b_{1}d_{1}h_{1}$ and $b_{2}d_{2}h_{2}$ are send to Bob. Then, all the photons should pass through the HWPs to perform the Hadamard operation. After the HWP operations, the states evolve to
\begin{eqnarray}
|\Gamma\rangle_{s_{1}}&\rightarrow&\frac{1}{4}\alpha(|H\rangle_{a_{1}}|H\rangle_{c_{1}}|H\rangle_{t_{1}}+|V\rangle_{a_{1}}|V\rangle_{c_{1}}|H\rangle_{t_{1}}+|H\rangle_{a_{1}}|V\rangle_{c_{1}}|V\rangle_{t_{1}}+|V\rangle_{a_{1}}|H\rangle_{c_{1}}|V\rangle_{t_{1}})\nonumber\\
&&(|H\rangle_{b_{1}}|H\rangle_{d_{1}}|H\rangle_{h_{1}}+|V\rangle_{b_{1}}|V\rangle_{d_{1}}|H\rangle_{h_{1}}+|H\rangle_{b_{1}}|V\rangle_{d_{1}}|V\rangle_{h_{1}}+|V\rangle_{b_{1}}|H\rangle_{d_{1}}|V\rangle_{h_{1}})\nonumber\\
&+&\frac{1}{4}\beta(|H\rangle_{a_{1}}|V\rangle_{c_{1}}|H\rangle_{t_{1}}+|V\rangle_{a_{1}}|H\rangle_{c_{1}}|H\rangle_{t_{1}}+|H\rangle_{a_{1}}|H\rangle_{c_{1}}|V\rangle_{t_{1}}+|V\rangle_{a_{1}}|V\rangle_{c_{1}}|V\rangle_{t_{1}})\nonumber\\
&&(|H\rangle_{b_{1}}|V\rangle_{d_{1}}|H\rangle_{h_{1}}+|V\rangle_{b_{1}}|H\rangle_{d_{1}}|H\rangle_{h_{1}}+|H\rangle_{b_{1}}|H\rangle_{d_{1}}|V\rangle_{h_{1}}+|V\rangle_{b_{1}}|V\rangle_{d_{1}}|V\rangle_{h_{1}}),\\
|T\rangle_{s_{2}}&\rightarrow&\frac{1}{4}\beta(|H\rangle_{a_{2}}|H\rangle_{c_{2}}|H\rangle_{t_{2}}+|V\rangle_{a_{2}}|V\rangle_{c_{2}}|H\rangle_{t_{2}}+|H\rangle_{a_{2}}|V\rangle_{c_{2}}|V\rangle_{t_{2}}+|V\rangle_{a_{2}}|H\rangle_{c_{2}}|V\rangle_{t_{2}})\nonumber\\
&&(|H\rangle_{b_{2}}|H\rangle_{d_{2}}|H\rangle_{h_{2}}+|V\rangle_{b_{2}}|V\rangle_{d_{2}}|H\rangle_{h_{2}}+|H\rangle_{b_{2}}|V\rangle_{d_{2}}|V\rangle_{h_{2}}+|V\rangle_{b_{2}}|H\rangle_{d_{2}}|V\rangle_{h_{2}})\nonumber\\
&+&\frac{1}{4}\alpha(|H\rangle_{a_{2}}|V\rangle_{c_{2}}|H\rangle_{t_{2}}+|V\rangle_{a_{2}}|H\rangle_{c_{2}}|H\rangle_{t_{2}}+|H\rangle_{a_{2}}|H\rangle_{c_{2}}|V\rangle_{t_{2}}+|V\rangle_{a_{2}}|V\rangle_{c_{2}}|V\rangle_{t_{2}})\nonumber\\
&&(|H\rangle_{b_{2}}|V\rangle_{d_{2}}|H\rangle_{h_{2}}+|V\rangle_{b_{2}}|H\rangle_{d_{2}}|H\rangle_{h_{2}}+|H\rangle_{b_{2}}|H\rangle_{d_{2}}|V\rangle_{h_{2}}+|V\rangle_{b_{2}}|V\rangle_{d_{2}}|V\rangle_{h_{2}}).
\end{eqnarray}
The state of all the system can be written as
\begin{eqnarray}
|T\rangle_{s_{3}}&=&\frac{1}{16}\alpha^{2}[(|H\rangle_{a_{1}}|H\rangle_{c_{1}}|H\rangle_{t_{1}}+|V\rangle_{a_{1}}|V\rangle_{c_{1}}|H\rangle_{t_{1}}+|H\rangle_{a_{1}}|V\rangle_{c_{1}}|V\rangle_{t_{1}}+|V\rangle_{a_{1}}|H\rangle_{c_{1}}|V\rangle_{t_{1}})\nonumber\\
&&(|H\rangle_{b_{1}}|H\rangle_{d_{1}}|H\rangle_{h_{1}}+|V\rangle_{b_{1}}|V\rangle_{d_{1}}|H\rangle_{h_{1}}+|H\rangle_{b_{1}}|V\rangle_{d_{1}}|V\rangle_{h_{1}}+|V\rangle_{b_{1}}|H\rangle_{d_{1}}|V\rangle_{h_{1}})\nonumber\\
&&(|H\rangle_{a_{2}}|V\rangle_{c_{2}}|H\rangle_{t_{2}}+|V\rangle_{a_{2}}|H\rangle_{c_{2}}|H\rangle_{t_{2}}+|H\rangle_{a_{2}}|H\rangle_{c_{2}}|V\rangle_{t_{2}}+|V\rangle_{a_{2}}|V\rangle_{c_{2}}|V\rangle_{t_{2}})\nonumber\\
&&(|H\rangle_{b_{2}}|V\rangle_{d_{2}}|H\rangle_{h_{2}}+|V\rangle_{b_{2}}|H\rangle_{d_{2}}|H\rangle_{h_{2}}+|H\rangle_{b_{2}}|H\rangle_{d_{2}}|V\rangle_{h_{2}}+|V\rangle_{b_{2}}|V\rangle_{d_{2}}|V\rangle_{h_{2}})]\nonumber\\
&+&\frac{1}{16}\beta^{2}(|H\rangle_{a_{1}}|V\rangle_{c_{1}}|H\rangle_{t_{1}}+|V\rangle_{a_{1}}|H\rangle_{c_{1}}|H\rangle_{t_{1}}+|H\rangle_{a_{1}}|H\rangle_{c_{1}}|V\rangle_{t_{1}}+|V\rangle_{a_{1}}|V\rangle_{c_{1}}|V\rangle_{t_{1}})\nonumber\\
&&(|H\rangle_{b_{1}}|V\rangle_{d_{1}}|H\rangle_{h_{1}}+|V\rangle_{b_{1}}|H\rangle_{d_{1}}|H\rangle_{h_{1}}+|H\rangle_{b_{1}}|H\rangle_{d_{1}}|V\rangle_{h_{1}}+|V\rangle_{b_{1}}|V\rangle_{d_{1}}|V\rangle_{h_{1}})\nonumber\\
&&(|H\rangle_{a_{2}}|H\rangle_{c_{2}}|H\rangle_{t_{2}}+|V\rangle_{a_{2}}|V\rangle_{c_{2}}|H\rangle_{t_{2}}+|H\rangle_{a_{2}}|V\rangle_{c_{2}}|V\rangle_{t_{2}}+|V\rangle_{a_{2}}|H\rangle_{c_{2}}|V\rangle_{t_{2}})\nonumber\\
&&(|H\rangle_{b_{2}}|H\rangle_{d_{2}}|H\rangle_{h_{2}}+|V\rangle_{b_{2}}|V\rangle_{d_{2}}|H\rangle_{h_{2}}+|H\rangle_{b_{2}}|V\rangle_{d_{2}}|V\rangle_{h_{2}}+|V\rangle_{b_{2}}|H\rangle_{d_{2}}|V\rangle_{h_{2}})\nonumber\\
&+&\frac{1}{16}\alpha\beta[(|H\rangle_{a_{1}}|H\rangle_{c_{1}}|H\rangle_{t_{1}}+|V\rangle_{a_{1}}|V\rangle_{c_{1}}|H\rangle_{t_{1}}+|H\rangle_{a_{1}}|V\rangle_{c_{1}}|V\rangle_{t_{1}}+|V\rangle_{a_{1}}|H\rangle_{c_{1}}|V\rangle_{t_{1}})\nonumber\\
&&(|H\rangle_{b_{1}}|H\rangle_{d_{1}}|H\rangle_{h_{1}}+|V\rangle_{b_{1}}|V\rangle_{d_{1}}|H\rangle_{h_{1}}+|H\rangle_{b_{1}}|V\rangle_{d_{1}}|V\rangle_{h_{1}}+|V\rangle_{b_{1}}|H\rangle_{d_{1}}|V\rangle_{h_{1}})\nonumber\\
&&(|H\rangle_{a_{2}}|H\rangle_{c_{2}}|H\rangle_{t_{2}}+|V\rangle_{a_{2}}|V\rangle_{c_{2}}|H\rangle_{t_{2}}+|H\rangle_{a_{2}}|V\rangle_{c_{2}}|V\rangle_{t_{2}}+|V\rangle_{a_{2}}|H\rangle_{c_{2}}|V\rangle_{t_{2}})\nonumber\\
&&(|H\rangle_{b_{2}}|H\rangle_{d_{2}}|H\rangle_{h_{2}}+|V\rangle_{b_{2}}|V\rangle_{d_{2}}|H\rangle_{h_{2}}+|H\rangle_{b_{2}}|V\rangle_{d_{2}}|V\rangle_{h_{2}}+|V\rangle_{b_{2}}|H\rangle_{d_{2}}|V\rangle_{h_{2}})\nonumber\\
&+&(|H\rangle_{a_{1}}|V\rangle_{c_{1}}|H\rangle_{t_{1}}+|V\rangle_{a_{1}}|H\rangle_{c_{1}}|H\rangle_{t_{1}}+|H\rangle_{a_{1}}|H\rangle_{c_{1}}|V\rangle_{t_{1}}+|V\rangle_{a_{1}}|V\rangle_{c_{1}}|V\rangle_{t_{1}})\nonumber\\
&&(|H\rangle_{b_{1}}|V\rangle_{d_{1}}|H\rangle_{h_{1}}+|V\rangle_{b_{1}}|H\rangle_{d_{1}}|H\rangle_{h_{1}}+|H\rangle_{b_{1}}|H\rangle_{d_{1}}|V\rangle_{h_{1}}+|V\rangle_{b_{1}}|V\rangle_{d_{1}}|V\rangle_{h_{1}})\nonumber\\
&&(|H\rangle_{a_{2}}|V\rangle_{c_{2}}|H\rangle_{t_{2}}+|V\rangle_{a_{2}}|H\rangle_{c_{2}}|H\rangle_{t_{2}}+|H\rangle_{a_{2}}|H\rangle_{c_{2}}|V\rangle_{t_{2}}+|V\rangle_{a_{2}}|V\rangle_{c_{2}}|V\rangle_{t_{2}})\nonumber\\
&&(|H\rangle_{b_{2}}|V\rangle_{d_{2}}|H\rangle_{h_{2}}+|V\rangle_{b_{2}}|H\rangle_{d_{2}}|H\rangle_{h_{2}}+|H\rangle_{b_{2}}|H\rangle_{d_{2}}|V\rangle_{h_{2}}+|V\rangle_{b_{2}}|V\rangle_{d_{2}}|V\rangle_{h_{2}}).]\label{evolve2}
\end{eqnarray}

Subsequently, as shown in Fig. \ref{f2}, the parties Alice and Bob let the photons $a_{1}$ and $a_{2}$, $c_{1}$ and $c_{2}$, $t_{1}$ and $t_{2}$, $b_{1}$ and $b_{2}$, $d_{1}$ and $d_{2}$, $h_{1}$ and $h_{2}$ pass through the six PBSs. They choose the case that all the output modes of the PBSs exactly contain one photon. The state in Eq. (\ref{evolve2}) becomes
\begin{eqnarray}
\rightarrow\frac{1}{4\sqrt{2}}(|H\rangle_{a_{1}}|H\rangle_{c_{1}}|H\rangle_{t_{1}}|H\rangle_{a_{2}}|H\rangle_{c_{2}}|H\rangle_{t_{2}}
+|V\rangle_{a_{1}}|V\rangle_{c_{1}}|H\rangle_{t_{1}}|V\rangle_{a_{2}}|V\rangle_{c_{2}}|H\rangle_{t_{2}}\nonumber\\
+|H\rangle_{a_{1}}|V\rangle_{c_{1}}|V\rangle_{t_{1}}|H\rangle_{a_{2}}|V\rangle_{c_{2}}|V\rangle_{t_{2}}
+|V\rangle_{a_{1}}|H\rangle_{c_{1}}|V\rangle_{t_{1}}|V\rangle_{a_{2}}|H\rangle_{c_{2}}|V\rangle_{t_{2}})\nonumber\\
(|H\rangle_{b_{1}}|H\rangle_{d_{1}}|H\rangle_{h_{1}}|H\rangle_{b_{2}}|H\rangle_{d_{2}}|H\rangle_{h_{2}}
+|V\rangle_{b_{1}}|V\rangle_{d_{1}}|H\rangle_{h_{1}}|V\rangle_{b_{2}}|V\rangle_{d_{2}}|H\rangle_{h_{2}}\nonumber\\
+|H\rangle_{b_{1}}|V\rangle_{d_{1}}|V\rangle_{h_{1}}|H\rangle_{b_{2}}|V\rangle_{d_{2}}|V\rangle_{h_{2}}
+|V\rangle_{b_{1}}|H\rangle_{d_{1}}|V\rangle_{h_{1}}|V\rangle_{b_{2}}|H\rangle_{d_{2}}|V\rangle_{h_{2}})\nonumber\\
+(|H\rangle_{a_{1}}|V\rangle_{c_{1}}|H\rangle_{t_{1}}|H\rangle_{a_{2}}|V\rangle_{c_{2}}|H\rangle_{t_{2}}
+|V\rangle_{a_{1}}|H\rangle_{c_{1}}|H\rangle_{t_{1}}|V\rangle_{a_{1}}|H\rangle_{c_{1}}|H\rangle_{t_{1}}\nonumber\\
+|H\rangle_{a_{1}}|H\rangle_{c_{1}}|V\rangle_{t_{1}}|H\rangle_{a_{2}}|H\rangle_{c_{2}}|V\rangle_{t_{2}}
+|V\rangle_{a_{1}}|V\rangle_{c_{1}}|V\rangle_{t_{1}}|V\rangle_{a_{2}}|V\rangle_{c_{2}}|V\rangle_{t_{2}})\nonumber\\
(|H\rangle_{b_{1}}|V\rangle_{d_{1}}|H\rangle_{h_{1}}|H\rangle_{b_{2}}|V\rangle_{d_{2}}|H\rangle_{h_{2}}
+|V\rangle_{b_{1}}|H\rangle_{d_{1}}|H\rangle_{h_{1}}|V\rangle_{b_{2}}|H\rangle_{d_{2}}|H\rangle_{h_{2}}\nonumber\\
+|H\rangle_{b_{1}}|H\rangle_{d_{1}}|V\rangle_{h_{1}}|H\rangle_{b_{2}}|H\rangle_{d_{2}}|V\rangle_{h_{2}}
+|V\rangle_{b_{1}}|V\rangle_{d_{1}}|V\rangle_{h_{1}}|V\rangle_{b_{2}}|V\rangle_{d_{2}}|V\rangle_{h_{2}}).
\end{eqnarray}
 Finally, the photons in the spatial modes $a_{2}b_{2}c_{2}d_{2}t_{2}h_{2}$ are detected by the six photon detectors $D_{1}D_{2}D_{3}D_{4}D_{5}D_{6}$ in the base $|\pm\rangle$. If the measurement results of both Alice and Bob are the same, i.e., $|+\rangle|+\rangle|+\rangle$ or $|-\rangle|-\rangle|-\rangle$, they can obtain
\begin{eqnarray}
|T\rangle_{s_{4}}&=&\frac{1}{\sqrt{2}}[\frac{1}{2}(|H\rangle_{a_{1}}|H\rangle_{c_{1}}|H\rangle_{t_{1}}+|H\rangle_{a_{1}}|V\rangle_{c_{1}}|V\rangle_{t_{1}}+|V\rangle_{a_{1}}|H\rangle_{c_{1}}|V\rangle_{t_{1}}+|V\rangle_{a_{1}}|V\rangle_{c_{1}}|H\rangle_{t_{1}})\nonumber\\
&&\frac{1}{2}(|H\rangle_{b_{1}}|H\rangle_{d_{1}}|H\rangle_{h_{1}}+|H\rangle_{b_{1}}|V\rangle_{d_{1}}|V\rangle_{h_{1}}+|V\rangle_{b_{1}}|H\rangle_{d_{1}}|V\rangle_{h_{1}}+|V\rangle_{b_{1}}|V\rangle_{d_{1}}|H\rangle_{h_{1}})\nonumber\\
&+&\frac{1}{2}(|H\rangle_{a_{1}}|H\rangle_{c_{1}}|V\rangle_{t_{1}}+|H\rangle_{a_{1}}|V\rangle_{c_{1}}|H\rangle_{t_{1}}+|V\rangle_{a_{1}}|H\rangle_{c_{1}}|H\rangle_{t_{1}}+|V\rangle_{a_{1}}|V\rangle_{c_{1}}|V\rangle_{t_{1}})\nonumber\\
&&\frac{1}{2}(|H\rangle_{b_{1}}|H\rangle_{d_{1}}|V\rangle_{h_{1}}+|H\rangle_{b_{1}}|V\rangle_{d_{1}}|H\rangle_{h_{1}}+|V\rangle_{b_{1}}|H\rangle_{d_{1}}|H\rangle_{h_{1}}+|V\rangle_{b_{2}}|V\rangle_{d_{1}}|V\rangle_{h_{1}})].\label{df}
\end{eqnarray}
 On the other hand, there are other measurement results such as  $|+\rangle|+\rangle|-\rangle$,  $|+\rangle|-\rangle|-\rangle$, and so on.
 The remained state can also be transformed into the former of Eq. (\ref{df}) with the help of phase-file operation and bit-flip operation. Finally, they can perform the Hadamard operations in each photon to convert Eq. (\ref{df}) to the maximally entangled C-GHZ state as
\begin{eqnarray}
|T\rangle_{s_{5}}&=&\frac{1}{\sqrt{2}}[\frac{1}{\sqrt{2}}(|H\rangle_{a_{1}}|H\rangle_{c_{1}}|H\rangle_{t_{1}}+|V\rangle_{a_{1}}|V\rangle_{c_{1}}|V\rangle_{t_{1}})\frac{1}{\sqrt{2}}(|H\rangle_{b_{1}}|H\rangle_{d_{1}}|H\rangle_{h_{1}}+|V\rangle_{b_{1}}|V\rangle_{d_{1}}|V\rangle_{h_{1}})\nonumber\\
&+&\frac{1}{\sqrt{2}}(|H\rangle_{a_{1}}|H\rangle_{c_{1}}|H\rangle_{t_{1}}-|V\rangle_{a_{1}}|V\rangle_{c_{1}}|V\rangle_{t_{1}})\frac{1}{\sqrt{1}}(|H\rangle_{b_{1}}|H\rangle_{d_{1}}|H\rangle_{h_{1}}-|V\rangle_{b_{1}}|V\rangle_{d_{1}}|V\rangle_{h_{1}})].\label{ds}
\end{eqnarray}
In this way, they can obtain the maximally C-GHZ state with the total success probability of $\frac{1}{8}|\alpha\beta|^{2}$.

\begin{figure}[!h]
\begin{center}
\includegraphics[width=10cm,angle=0]{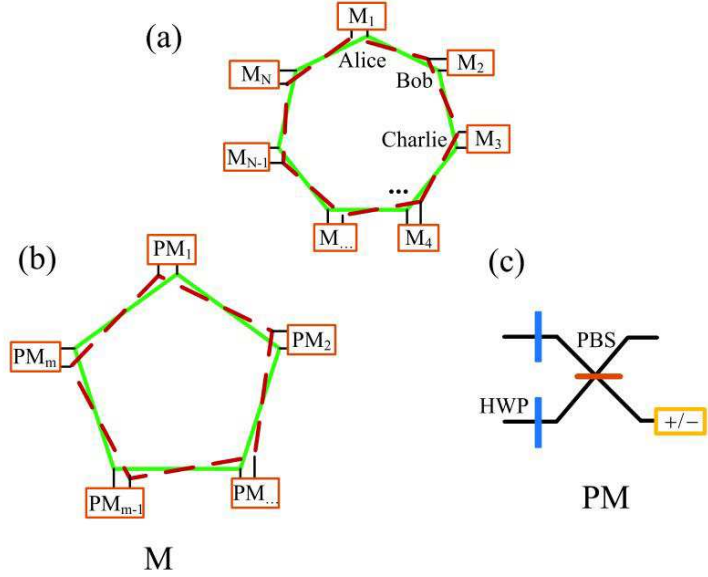}
\caption{(a) The schematic diagram of ECP for arbitrary C-GHZ state. Red dash and green solid denote two pairs less-entangled C-GHZ state, respectively, which contains N logic qubits. Two logical qubit in one place make the measurement. (b) The schematic diagram of two logical qubit making the PM measurement. (c) The schematic diagram of PM measurement.}\label{f3}
\end{center}
\end{figure}

\section{The ECP for arbitrary C-GHZ state}

Finally, let us briefly describe our  ECP for the arbitrary C-GHZ state. The less-entangled C-GHZ state are shared by $N$ parties (Alice, Bob, Charlie, ect). Each one owns one logic qubit. As shown in Fig. \ref{f3}, two pairs of less-entangled C-GHZ states are sent to the $N$ parties, which can be written as
\begin{eqnarray}
&&|\Psi_{1}\rangle_{N,m}=\alpha|GHZ^{+}_{m}\rangle^{\otimes N}+\beta|GHZ^{-}_{m}\rangle^{\otimes N},\label{q18}\\
&&|\Psi_{2}\rangle'_{N,m}=\alpha|GHZ^{+}_{m}\rangle^{\otimes N}+\beta|GHZ^{-}_{m}\rangle^{\otimes N}.
\end{eqnarray}

Before concentration, by using the Hadamard operation and single qubit rotation, which is the same as the previous description, the state $|\Psi_{2}\rangle'_{N,m}$ can be transformed into
\begin{eqnarray}
|\Psi_{2}\rangle_{N,m}=\beta|GHZ^{+}_{m}\rangle^{\otimes N}+\alpha|GHZ^{-}_{m}\rangle^{\otimes N}.
\end{eqnarray}
Firstly, all the parties pass through the HWPs. The states will evolve to
\begin{eqnarray}
|\Psi_{1}\rangle_{N,m}&\rightarrow&\alpha[(\frac{1}{\sqrt{2}}(|H\rangle+|V\rangle))^{\otimes m}+(\frac{1}{\sqrt{2}}(|H\rangle-|V\rangle))^{\otimes m}]^{\otimes N}\nonumber\\
&&+\beta[(\frac{1}{\sqrt{2}}(|H\rangle+|V\rangle))^{\otimes m}-(\frac{1}{\sqrt{2}}(|H\rangle-|V\rangle))^{\otimes m}]^{\otimes N},\\
|\Psi_{2}\rangle_{N,m}&\rightarrow&\beta[(\frac{1}{\sqrt{2}}(|H\rangle+|V\rangle))^{\otimes m}+(\frac{1}{\sqrt{2}}(|H\rangle-|V\rangle))^{\otimes m}]^{\otimes N}\nonumber\\
&&+\alpha[(\frac{1}{\sqrt{2}}(|H\rangle+|V\rangle))^{\otimes m}-(\frac{1}{\sqrt{2}}(|H\rangle-|V\rangle))^{\otimes m}]^{\otimes N}.
\end{eqnarray}
The whole state can be written as
\begin{eqnarray}
|\Psi_{s}\rangle_{N,m}&=&\alpha^{2}[(\frac{1}{\sqrt{2}}(|H\rangle+|V\rangle))^{\otimes m}+(\frac{1}{\sqrt{2}}(|H\rangle-|V\rangle))^{\otimes m}]^{\otimes N}\nonumber\\
&&[(\frac{1}{\sqrt{2}}(|H\rangle+|V\rangle))^{\otimes m}-(\frac{1}{\sqrt{2}}(|H\rangle-|V\rangle))^{\otimes m}]^{\otimes N}\nonumber\\
&+&\beta^{2}[(\frac{1}{\sqrt{2}}(|H\rangle+|V\rangle))^{\otimes m}-(\frac{1}{\sqrt{2}}(|H\rangle-|V\rangle))^{\otimes m}]^{\otimes N}\nonumber\\
&&[(\frac{1}{\sqrt{2}}(|H\rangle+|V\rangle))^{\otimes m}+(\frac{1}{\sqrt{2}}(|H\rangle-|V\rangle))^{\otimes m}]^{\otimes N}\nonumber\\
&+&\alpha\beta\{[(\frac{1}{\sqrt{2}}(|H\rangle+|V\rangle))^{\otimes m}+(\frac{1}{\sqrt{2}}(|H\rangle-|V\rangle))^{\otimes m}]^{\otimes N}\nonumber\\
&&[(\frac{1}{\sqrt{2}}(|H\rangle+|V\rangle))^{\otimes m}+(\frac{1}{\sqrt{2}}(|H\rangle-|V\rangle))^{\otimes m}]^{\otimes N}\nonumber\\
&&+[(\frac{1}{\sqrt{2}}(|H\rangle+|V\rangle))^{\otimes m}-(\frac{1}{\sqrt{2}}(|H\rangle-|V\rangle))^{\otimes m}]^{\otimes N}\nonumber\\
&&[(\frac{1}{\sqrt{2}}(|H\rangle+|V\rangle))^{\otimes m}-(\frac{1}{\sqrt{2}}(|H\rangle-|V\rangle))^{\otimes m}]^{\otimes N}\}.\label{mn}
\end{eqnarray}
 In Fig. \ref{f3}, figure (a) means that the two pairs of C-GHZ states are shared by each parity, and each parity will perform the measurement on his or her two logic qubits. Figure (b) means that each party owns two logic qubits, which are encoded in physical polarized GHZ states, respectively. The PM measurement performed on the GHZ states in each party is analogy with Fig. \ref{f2}. Each party should perform  PM measurements for $m$ times. Figure (c) is the schematic of the PM measurement. if all the output modes of PBS in each PM measurement contain only one photon,  the ECP for the arbitrary C-GHZ is successful.   Finally, they measure all the photons in the lower output modes of each PBS, if the measurement result is $|\pm\rangle^{\otimes mN}$, they will obtain the maximally entangled state
\begin{eqnarray}
|\Psi_{so}\rangle_{N,m}=\frac{1}{\sqrt{2}}(|GHZ^{+}_{m}\rangle^{\otimes N}+|GHZ^{-}_{m}\rangle^{\otimes N}).
\end{eqnarray}
The total success probability is
\begin{eqnarray}
P_{mN}=\frac{|\alpha\beta|^{2}}{2^{(m-1)N-1}}.
\end{eqnarray}

\begin{figure}[!h]
\begin{center}
\includegraphics[width=12cm,angle=0]{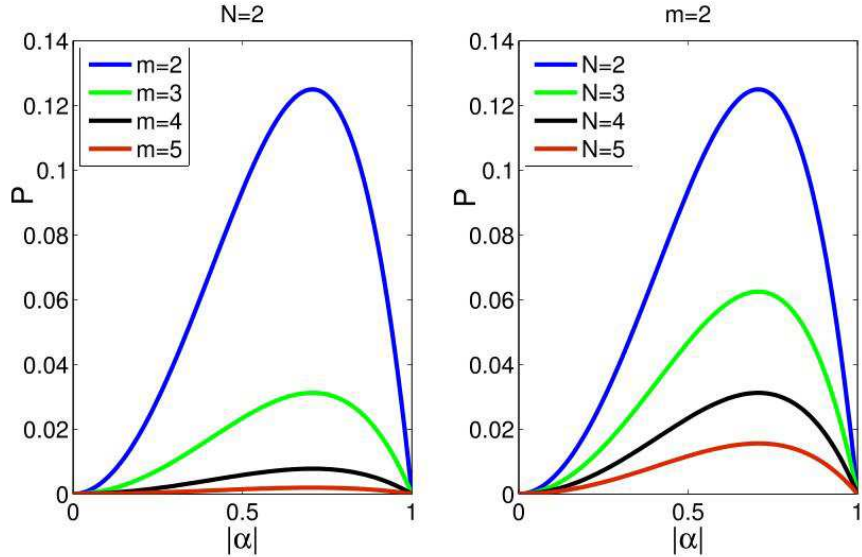}
\caption{The success probability of the ECP for the arbitrary C-GHZ state with the different values of $m$ and $N$.}\label{fmN}
\end{center}
\end{figure}
\section{Discussion and conclusion}
So far, we have completely described our ECP. We first described the ECP with a simple case that $m=N=2$. Subsequently, we explain this protocol for
concentrating the C-GHZ state with $m=3$, $N=2$, and arbitrary $m$, $N$, respectively. In Fig. 4, we calculated the success probability $P_{mN}$ altered with $m$ and $N$, respectively. We show that the larger of $m$ and $N$, the smaller of  $P_{mN}$. In the previous works of entanglement concentration, they exploited the controlled-not (CNOT) gate and nonlinear cross-Kerr nonlinearity to complete the task \cite{Qu,panjun}. If consider the practical experiment condition, both the CNOT gate and
cross-Kerr nonlinearity are hard to realize. In this ECP, we only require the linear optics, such as PBSs, HWPs   to complete the task. Moreover, this ECP does not know the exact information of the initial state. These advantages make this protocol feasible in current QIP tasks.

\section*{ACKNOWLEDGEMENTS}
This work was supported by the National Natural Science Foundation
of China under Grant  Nos. 11474168 and 61401222, the Natural Science Foundation of Jiangsu province under Grant No. BK20151502,
the Qing Lan Project in Jiangsu Province, and a Project
Funded by the Priority Academic Program Development of Jiangsu
Higher Education Institutions.\\

\end{document}